\documentclass[aps,prx,floats,twocolumn,superscriptaddress]{revtex4-1}
\usepackage{rotating}
\usepackage{graphicx,epsfig}
\usepackage{graphics,dcolumn,bm,epic, eepic,fleqn,float}
\usepackage{amssymb,amsmath,amsfonts,multirow,rotate,color}
\usepackage{soul}
\usepackage{wasysym}
\usepackage{xcolor}
\usepackage{xkeyval}
\usepackage{adjustbox}
\usepackage{array}

\newcommand{\DCI}{DCI~}
\newcommand{\DCIz}{DCI$_z\>$}

\newdimen\mycfsdim
\newcommand{\mycfs}[1]{\mycfsdim=#1pt 
\mycfsdim=1.2\mycfsdim 
	\fontsize{#1pt}{\the\mycfsdim}\selectfont}

\newcolumntype{R}[2]{%
	>{\adjustbox{angle=#1,lap=\width-(#2)}\bgroup}%
	l%
	<{\egroup}%
}
\newcommand*\rot{\multicolumn{1}{R{45}{1em}}}

\newcommand{\lay}[1]{^{[#1]}}

\usepackage{color}
\usepackage{lipsum}
\usepackage{mathrsfs}

\definecolor{darkblue}{rgb}{0,0,0.555}
\usepackage[colorlinks=true,
            linkcolor=black,
            urlcolor=blue,
            citecolor=blue,
            bookmarks=true,
            pdfborder={0 0 0}]{hyperref}     
\newcommand{\newtext}[1]{\leavevmode\textcolor{black}{#1}}
\newcommand{\Erdos}{Erd\"{o}s}
\newcommand{\Renyi}{R\'{e}nyi }
\begin{document}

\title{Optimal percolation in correlated multilayer networks with
  overlap}

\author{Andrea Santoro}
\affiliation{School of Mathematical Sciences, Queen Mary University of
	London, Mile End Road, E1 4NS, London (UK)}
\affiliation{The Alan Turing Institute, The British Library, NW1 2DB,
	London, United Kingdom}
\author{Vincenzo Nicosia}
\affiliation{School of Mathematical Sciences, Queen Mary University of
	London, Mile End Road, E1 4NS, London (UK)}

\begin{abstract}
  \newtext{Multilayer networks have been found to be prone to abrupt
    cascading failures under random and targeted attacks, but most of
    the targeting algorithms proposed so far have been mainly tested
    on uncorrelated systems. Here we show that the size of the
    critical percolation set of a multilayer network is substantially
    affected by the presence of inter-layer degree correlations and
    edge overlap. We provide extensive numerical evidence which
    confirms that the state-of-the-art optimal percolation strategies
    consistently fail to identify minimal percolation sets in
    synthetic and real-world correlated multilayer networks, thus
    overestimating their robustness. We propose two new targeting
    algorithms, based on the local estimation of path disruptions away
    from a given node, and a family of Pareto-efficient strategies
    that take into account both intra-layer and inter-layer
    heuristics, and can be easily extended to multiplex networks with
    an arbitrary number of layers. We show that these strategies
    consistently outperform existing attacking algorithms, on both
    synthetic and real-world multiplex networks, and provide some
    interesting insights about the interplay of correlations and
    overlap in determining the hyperfragility of real-world multilayer
    networks. Overall, the results presented in the paper suggest that
    we are still far from having fully identified the salient
    ingredients determining the robustness of multiplex networks to
    targeted attacks.}
\end{abstract}

\maketitle

\section{Introduction}
Network percolation theory has been recently shaken by the discovery
that interdependencies and feedback loops between interacting networks
change the character of the percolation transition and make it
explosive~\cite{Buldyrev_Havlin2010,Boccaletti2016explosive,DSouza2019explosive}.
These results have acquired even more relevance in the last few years,
due to the increasing experimental evidence about real-world systems
whose structures are naturally represented as
multiplex~\cite{Arenas_2013,Boccaletti_2014} or multilayer
networks~\cite{Bianconi2018multilayer}. Random percolation in
multiplex networks is nowadays quite
well-understood~\cite{gao2012networks,cellai2013percolation,radicchi2015percolation,Bianconi2016percolation,Coghi2018controlling},
and the wide spectrum of possible percolation transitions, from abrupt
to continuous
ones~\cite{Baxter2012avalanche,baxter2016correlated,Makse_NatComm_correlation_2014},
has been successfully related to some of the structural properties of
these networks, such as the presence of inter-layer degree correlations
and edge
overlap~\cite{Min2014network,Nicosia2015,cellai2016message,Bianconi2016percolation,Kryven_Bianconi2019}.
However, quite often targeted attacks can potentially drive a system
to collapse by knocking down a much smaller fraction of nodes than
required by random
attacks~\cite{Barabasi2000,Callaway2000,Dong2013robustness_targeted,Zhao2016robustness_targeted}. Hence,
optimal percolation, that is the problem of finding the minimal
fraction of nodes whose removal would irreversibly fragment the
system, has been extensively studied both in single-layer
networks~\cite{Makse_review_2019} and, more recently, in multilayer
networks as well~\cite{osat2017optimal}.
\newtext{One of the reasons behind this renovated interest for optimal
  percolation is the fact that targeted attacks also play a central
  role in optimal strategies for influence maximisation in opinion
  dynamics~\cite{Kempe2003maximizing, Kitsak2010identification} and
  for effective immunisation in spreading
  processes~\cite{Cohen2003efficient, Pastor2002immunization,
    clusella2016immunization}}. The fact that most of the single-layer
optimal attack strategies~\cite{Morone2015influence,
  zdeborova2016fast, mugisha2016identifying, clusella2016immunization}
cannot be easily extended to the multilayer case, has resulted in an
interesting and quite active line of
research~\cite{osat2017optimal,Baxter2018targeted}. Although
correlations and overlap are indeed a salient aspect of all real-world
multiplex
networks~\cite{Bianconi2018multilayer,Nicosia2015,Min2014network,kleineberg2017geometric},
the few strategies for optimal multiplex percolation proposed so far
have been mainly tested on synthetic uncorrelated multilayer networks,
thus neglecting inter-layer degree correlations and edge overlap.

\newtext{In this work we fill this gap by studying the problem of
  optimal percolation in multilayer networks with non-trivial
  inter-layer degree correlations and non-negligible edge overlap. We
  find that the robustness of systems under targeted attacks is deeply
  affected by the presence of both inter-layer degree correlations and
  edge overlap. In particular, all the current algorithms for optimal
  percolation systematically overestimate the size $q$ of the minimal
  set of nodes to knock down in order to destroy the mutually
  connected giant component. Here we introduce two new classes of
  algorithms, respectively based on a generalisation to duplex
  networks of the Collective Influence
  algorithm~\cite{Morone2015influence}, and on the concept of
  Pareto-efficiency~\cite{Deb2001_MOO,Miettinen2012_MOO}, which allows
  to combine layer-based and genuinely multi-layer node properties. We
  show through extensive numerical simulations that all these
  algorithms provide consistently smaller critical sets in synthetic
  correlated multilayer networks, and outperform other
  state-of-the-art algorithms in real-world systems.}

\section{Targeted attack strategies}
Let us consider a multiplex network $\mathcal{M}$ with $N$ nodes and
two layers. The undirected and unweighted edges on each layer are
encoded in the adjacency matrices $A_{ij}^{[\alpha]},\> \alpha=1,2$,
whose generic element $A_{ij}^{[\alpha]} = 1$ if and only if nodes $i$
and $j$ are the endpoints of an edge at layer $\alpha$, \newtext{and
  is zero otherwise}. Two nodes of $\mathcal{M}$ belong to the same
Mutually Connected Component (MCC) if there exists at least one path
on each layer that connects them and traverses only nodes belonging to
the same MCC. \newtext{The parameter of interest in percolation
  analysis is the relative size of the largest mutually connected
  component (LMCC), that is the largest maximal sub-graph consisting of
  mutually connected
  nodes~\cite{Buldyrev_Havlin2010,Bianconi2018multilayer}. Notice that
  the LMCC is a generalisation of the giant connected component for
  single-layer graphs. The optimal percolation problem consists in
  finding the smallest set of nodes which, if removed, would reduce
  the size of the LMCC to $O(N^{1/2})$. We call this set
  \textit{critical set} or \textit{attack set}, and we denote its
  relative size with $q$.}

\newtext{Optimal percolation is naturally a many-body problem. Indeed,
  interactions among nodes at all distances play an important role in
  the determination of the damage caused by the removal of a subset of
  nodes, which makes the problem
  NP-hard~\cite{Kempe2003maximizing}. Although there are currently no
  studies about the computational complexity of optimal multiplex
  percolation, it is reasonable to assume that this problem is not
  easier than it's classical single-layer counterpart, especially
  because the computation of the LMCC is based on the existence of
  paths connecting each pair of nodes on two graphs at the same
  time. Hence the necessity to use heuristic algorithms to find
  approximate solutions.  In most of the cases, heuristic algorithms
  proceed by assigning a score to each node, based on some structural
  indicator~\cite{Makse_review_2019,newman2010networks,Latora_Nicosia_Russo_book2017},
  and then iteratively removing nodes in decreasing order of their
  score.}  As confirmed by recent
studies~\cite{osat2017optimal,Baxter2018targeted}, single-layer attack
strategies cannot be easily generalised to the case of multiplex
networks, mainly because it is not immediate to combine node scores on
different layers to obtain a meaningful ranking. The authors of
Ref.~\cite{osat2017optimal,Baxter2018targeted} proposed several ways
of integrating scores based on popular single-layer strategies,
namely: \textit{(i)} rankings based on the sum or product of the
degrees in the two layers (HDA)~\cite{osat2017optimal}; \textit{(ii)}
a generalisation of the Collective Influence Propagation
algorithm~\cite{Morone2016collective}; \textit{(iii)} a generalisation
of the so-called CoreHD algorithm (CoreHD)~\cite{zdeborova2016fast}.

\newtext{To give an idea of how hard it is to directly adapt a
  single-layer percolation strategy to a multi-layer setup, let us
  consider the CoreHD algorithm, which is one of the most effective
  strategies to destroy the giant connected component of a
  single-layer graph. The algorithm proceeds by iteratively removing
  the nodes with the highest degrees from the 2-core of the graph
  (i.e., by effectively de-cycling the network). However, this idea
  cannot be directly applied to duplex networks, since the 2-core of a
  multiplex graph is not uniquely defined. As a consequence, there are
  several existing multiplex extensions of the CoreHD strategy, but
  none of them provides satisfactory results on duplex
  networks~\cite{Baxter2018targeted}. By contrast, the
  recently-proposed Effective Multiplex Degree (EMD)
  strategy~\cite{Baxter2018targeted} consistently improves over all
  the other existing methods. Indeed, the heuristic used by EMD takes
  into account multi-layer adjacency at different distances, and
  effectively exploit the degree-heterogeneity between different
  layers.}

\subsection{Duplex Collective Influence}
\newtext{An efficient heuristic for optimal single-layer percolation
  was introduced in Ref.~\cite{Morone2015influence} by Morone and
  Makse. The authors mapped optimal percolation into the minimisation
  of energy of a many-body system, in which the interactions among
  units are expressed in terms of the non-backtracking matrix of the
  graph, and proposed an efficient and scalable algorithm, called
  Collective Influence (CI), to identify the minimal set of
  influential nodes to remove. The CI algorithm iteratively removes
  nodes according to the highest values of CI scores, defined as:
  \begin{equation}
    CI_\ell(i)=(k_i-1) \sum_{j \in \partial \mathcal{B}(i,\ell
      )}(k_j-1)
    \label{eq:CI}
  \end{equation}
  where $k_i$ is the degree of node $i$, while $\partial
  \mathcal{B}(i,\ell )$ represents the frontier of the ball of radius
  $\ell$ containing all the nodes at distance smaller than or equal to
  $\ell$ from node $i$. This means that a node $i$ is assigned a
  larger CI score if the set of nodes at distance $\ell$ from $i$ has
  a large number of links. By removing a node with a large CI score,
  we are potentially removing a node that mediates a large number of
  walks. Remarkably, the attack strategy based on CI can be
  implemented by an algorithm with time complexity $\mathcal{O}(N \log
  N)$, which is attained by using a max-heap to keep and update the CI
  scores of nodes~\cite{Morone2016collective}. Some variations of the
  CI heuristic have managed to obtain relatively better performance
  (i.e., smaller attack sets) by including more structural information
  about the relevance of a given node for
  percolation~\cite{Morone2016collective}, and at the cost of an
  increased time complexity. There has also been an attempt to extend
  the Collective Influence algorithm to the case of duplex networks by
  combining the bare CI scores of the nodes at the two
  layers~\cite{Baxter2018targeted}, but the results are not
  competitive with other existing algorithms. The main reason is that
  the bare combination of the layer-based scores does not take into
  account the role played by edge overlap and inter-layer degree
  correlations in triggering a cascade of node removals.}

\newtext{We introduce here two generalisations of Collective Influence
  for duplex networks, which automatically take into account both
  inter-layer degree correlations and edge overlap. The heuristics are
  based on two simple ideas: the first one is that nodes with high
  degrees and high edge overlap are more likely responsible for
  mediating a lot of interdependent paths; the second one is that the
  removal of a given node $i$ has a large impact on the size of the
  MCC if it triggers a larger cascade of node removals \textit{away}
  from $i$. We define the Duplex Collective Influence (\DCI) as
  follows:
  \begin{equation}
    DCI(i)= \frac{k_i\lay{1} k_i\lay{2}-k_i^{\rm int}}{k_i^{\rm aggr}}
    \left[\sum_{j} a_{ij}\lay{1} (k_j\lay{2} - 1) +
      a_{ij}\lay{2}(k_j\lay{1} - 1)\right]
    \label{eq:DCI_formula}
  \end{equation}
  where $k_i^{\rm int}$ is the degree of node $i$ in the intersection
  graph (i.e., the graph containing only the links which appear in
  both layers), and $k_i^{\rm aggr}$ is the degree of node $i$ in the
  binary aggregated graph (i.e., the union graph obtained by
  collapsing the two layers into one~\cite{battiston2014}). The \DCI
  score of a given node $i$ is indeed obtained as the product of two
  terms. The first contribution is due to the product of the degrees
  of node $i$ at the two layers and to the local edge overlap of node
  $i$. It is easy to show that this term increases when $k_i^{\rm int}$
  increases, meaning that nodes with a high edge overlap and high
  degrees a the two layers will in general be ranked higher (see
  Appendix~\ref{appendix:DCI} for additional details).  The term in
  square brackets, instead, takes into account potential cascades away
  from node $i$ triggered by the removal of $i$. In particular, the
  term is larger if the neighbours of $i$ on layer 1 have a high
  degree on layer 2, and vice-versa. In this case, the removal of $i$
  (and of all its edges on both layers) will disrupt all the paths
  between the neighbours of $i$ on layer 2 which are mediated by $i$,
  hence potentially disrupting the connected component to which $i$
  belongs at layer $2$. This might in turn trigger further node
  removals in the neighbourhoods of those nodes, and let the cascade
  propagate away from node $i$. In the limiting case of a duplex
  network consisting of two identical layers (which is indeed
  equivalent to a single-layer network with respect to percolation),
  \DCI yields the same node ranking as that induced by CI on the
  aggregated network when we set $\ell=1$ in Eq.(\ref{eq:CI}) [see
  Appendix~\ref{appendix:DCI} for details].}

\newtext{It is important to note that when nodes are iteratively
  removed from a duplex, the term $k_i\lay{1} k_i\lay{2}$ in
  Eq.~(\ref{eq:DCI_formula}) might become equal to zero, e.g., due to
  the removal of nodes around $i$ which have left node $i$ isolated in
  one of the two layers. However, node $i$ might still have a
  relatively large degree on the other layer, and its removal might
  trigger larger cascades away from $i$ than a node which is still
  connected on both layers but has a small degree on each of
  them. This happens more frequently in duplex networks with
  heterogeneous degree distributions. To account for this
  inconvenience, we define a modified \DCI score:
  \begin{equation}
    \begin{aligned}
    \small DCI_{z}(i)
    &
    =\frac{(k_i\lay{1}+1)(k_i\lay{2}+1)-3k_i^{int}-1}{k_i^{aggr}}
    \times\\
    & \times
    \left[\sum_{j} a_{ij}\lay{1} (k_j\lay{2} - 1) +
      a_{ij}\lay{2}(k_j\lay{1} - 1)\right]
    \end{aligned}
    \label{eq:DCI_z}
  \end{equation}
  which is obtained by replacing $k_i\lay{\alpha}$ with
  $k_i\lay{\alpha} + 1$ in Eq.~(\ref{eq:DCI_formula}), and enforcing
  that \DCIz induces the same node ranking as CI with $\ell = 1$ in
  the limiting case of duplex networks made of two identical layers
  (see Appendix~\ref{appendix:DCI} for details). The subscript $z$
  indicates that we are correcting for nodes with zero degree on at
  least one of the two layers.}

\newtext{We use \DCI and \DCIz in an adaptive algorithm that
  iteratively removes nodes from the duplex according to their score
  recomputed on the remaining sub-graph.  This process is iterated
  until the size of the LMCC becomes non-extensive
  [i.e. $\mathcal{O}(N^{1/2}$)]. The time complexity of the direct
  implementation of this algorithm by means of simple data structures
  is $O(N^2\log N)$, but a more efficient algorithm which uses a max-heap
  to keep the list of scores sorted will have time complexity
  $O(N^{1.2})$ (see Appendix~\ref{appendix:time_complexity} for
  details)}.

\subsection{Pareto-efficiency for multi-objective optimisation}

\begin{figure}[!t]
	\includegraphics[width=3in]{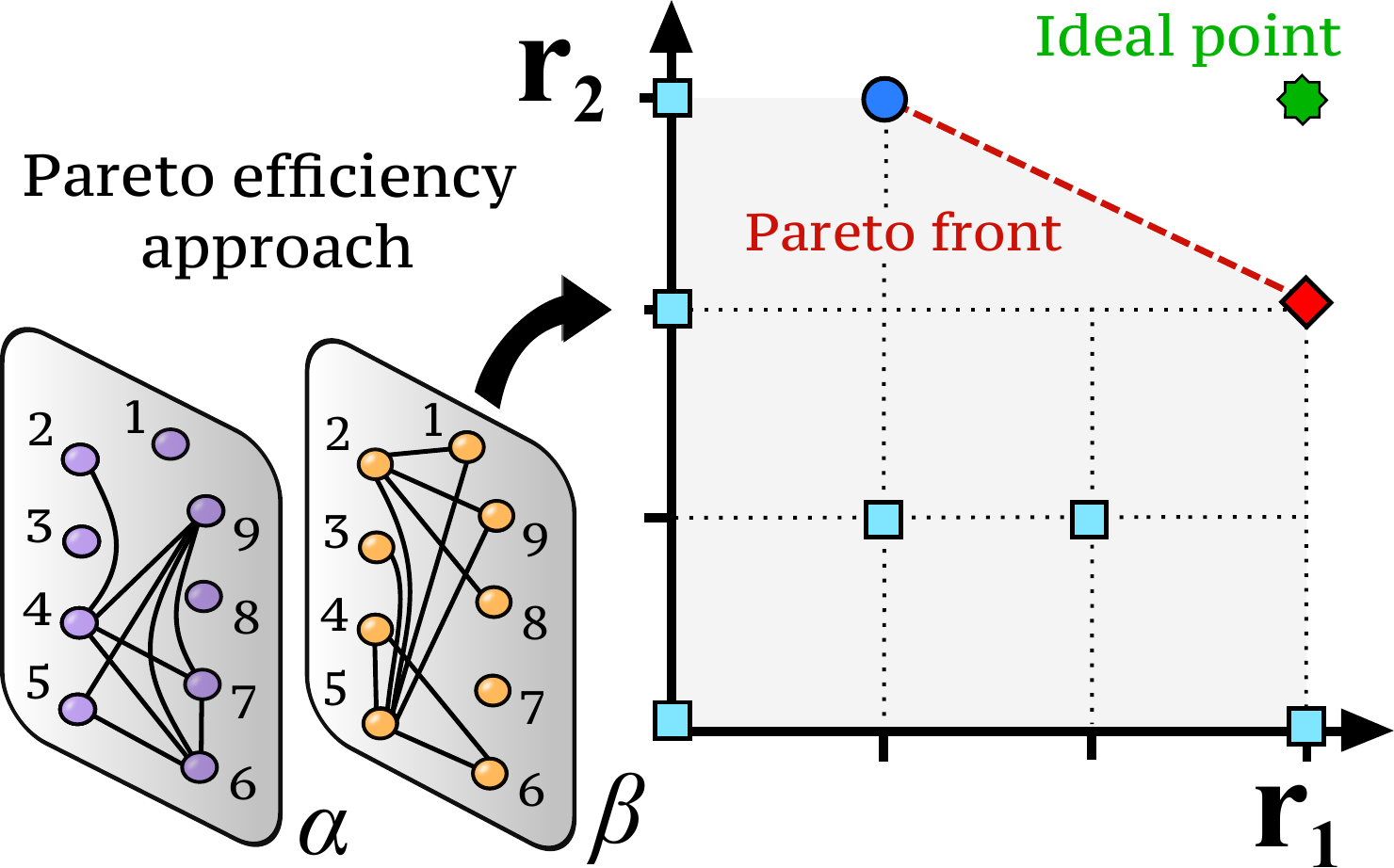}
	\caption{Graphical representation of Pareto efficiency for two
    generic structural node descriptors $\bm{r_1}$ and
    $\bm{r_2}$. Each node of the multiplex is mapped onto a point in
    the $(\bm{r_1}, \bm{r_2})$ plane. The points for which no
    improvement can be achieved in one objective function without
    hindering the others are called Pareto-efficient (blue circle and
    red diamond) and constitute a Pareto front. The Pareto-efficient
    points are iteratively ranked according to their Euclidean
    distance from the ideal point (green star), i.e., the point that
    maximises all the objective functions. In this case, the node
    associated to the red diamond is ranked first.}
	\label{fig:fig0}
\end{figure}

\newtext{A second class of attack strategies is based on the
  hypothesis} that it should be possible to obtain smaller attack sets
by combining layer-specific and genuinely multilayer information. We
use here the concept of Pareto
efficiency~\cite{Deb2001_MOO,Miettinen2012_MOO}, which was originally
devised to concurrently optimise multiple cost functions. The idea is
illustrated in Fig.~\ref{fig:fig0}. We consider a set of $m$ node
descriptors (also called objective functions), which we deem relevant
for multilayer percolation, so that each node $i$ is associated to the
vector of ranks induced by each of the $m$ scores
$\bm{r}^i=[r_{1}^i,r_{2}^i,\ldots,r_{m}^i]$, and is mapped into a
point of an $m$-dimensional space $C$. Assuming that optimal attack
sets consist of nodes who are maximising all the structural
descriptors at the same time, we can employ the concept of dominance
strict partial order~\cite{Miettinen2012_MOO} to identify
Pareto-efficient nodes in the space $C$. A point is considered
Pareto-efficient if no single score associated to node $i$ can be
improved without hindering the other scores associated to node $i$. In
general, for a given set of points there are more than one
Pareto-efficient points, which constitute the so-called Pareto front
for that set (see Fig.~\ref{fig:fig0}).
\begin{figure*}[!htbp]
	\includegraphics[width=0.95\textwidth]{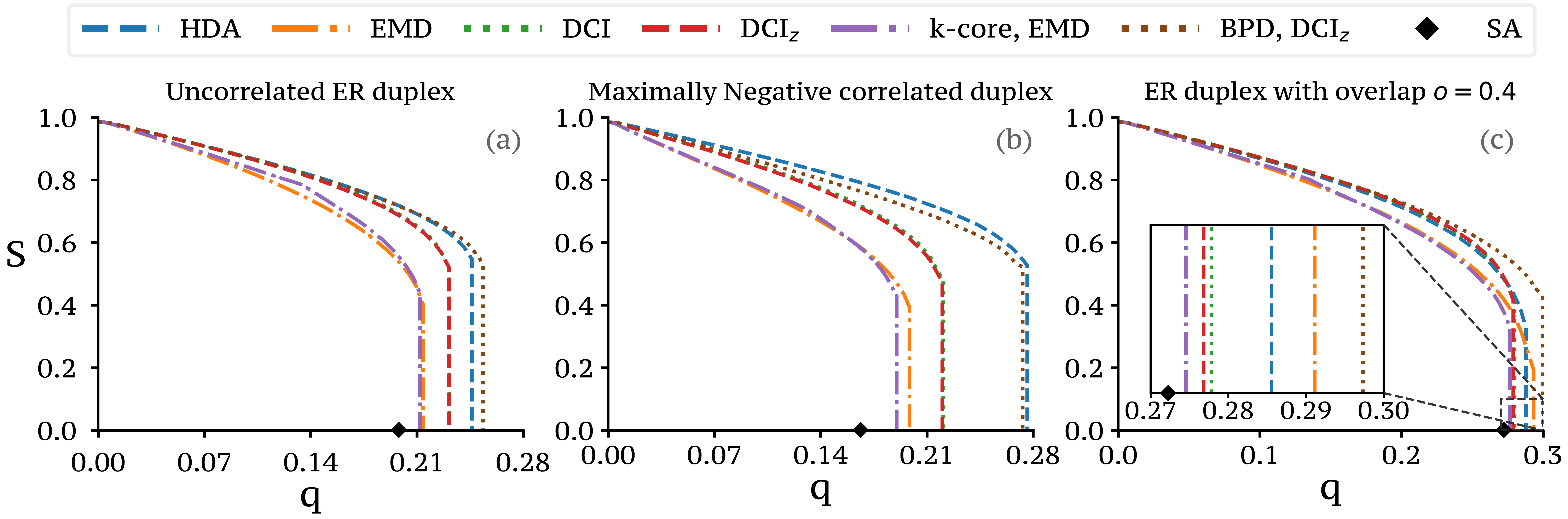}
	\caption{(a) \newtext{Relative size $S$ of the LMCC of a multiplex
      network when a fraction $q$ of nodes is removed by different
      targeted attack strategies. The multiplex consists of two
      \Erdos-\Renyi layers with $N=10^4$ nodes, average degree
      $\langle k \rangle = 5$, no inter-layer degree correlations
      ($\rho \approx 0$) and no edge overlap ($o \approx 0$). In this
      case, EMD provides a much smaller attack set than HDA, but the
      Pareto-efficient (k-core, EMD) strategy produces a smaller
      critical set.}  (b) If the two layers have maximally
    disassortative interlayer degree correlations ($\rho = - 1$) but
    still no edge overlap ($o\approx 0$), HDA performs sensibly worse
    than in the uncorrelated case, while the Pareto-efficient
    (k-core, EMD) finds a smaller attack set and again outperforms
    EMD. (c) If the duplex has a substantial edge overlap ($o=0.4$)
    and no correlations ($\rho \approx 0$), the critical set is much
    larger than in the other two cases. \newtext{The presence of edge
      overlap favours HDA, but the smallest attack set is still found
      by the (k-core, EMD) Pareto-efficient algorithm, immediately
      followed by \DCI and \DCIz. For comparison, we also report in
      each plot the results obtained by Simulated Annealing (black
      diamond). All the curves are averaged over 20 realisations.}}
	\label{fig:fig1}
\end{figure*}
\newtext{At a first glance, the Pareto-efficiency approach might
  appear similar to the hybrid methods presented in
  Ref.~\cite{Erkol2019hybrid}, however, there are a few fundamental
  differences. In particular, the Pareto-efficiency approach:
  \textit{i)} is agnostic with respect to the functions to be
  maximised (i.e., it is parameter--free); \textit{ii)} it has a
  simple physical interpretation (i.e., multi-objective optimisation
  arises naturally whenever a system is subject to at least two
  concurrent sets of constraints); and \textit{iii)} is known to have
  several advantages over scalarisation
  methods~\cite{Miettinen2012_MOO,Ehrgott2005multicriteria}.}

\newtext{Although multi-objective optimisation is a quite appealing
  concept, the main drawback is that it proposes a set of
  equally-viable ``optimal'' solutions at each iteration, and such a
  set normally contains multiple solutions. This is indeed far from
  ideal, since comparing the performance of different cost functions
  (obtained from different ways of ranking nodes on the basis of their
  structural properties) can become somehow complicated.  A common way
  to select only one of the Pareto-optimal solutions from a Pareto
  front, when no additional information is available about how
  preferable a certain solution is, consists in selecting the closest
  solution to the ideal point~\cite{Deb2001_MOO}, i.e. the (possibly
  non-existent) point that simultaneously maximises all the cost
  functions (see Fig.~\ref{fig:fig0}).  Alternative ways to select
  Pareto-optimal solutions exist in the
  literature~\cite{Miettinen2012_MOO,Ehrgott2005multicriteria,Keeney1993decisions},
  however, no consensus on the best approach has been reached yet. We
  adopted the ideal point method for the results shown in the
  following.  In other words, for each Pareto strategy, we constructed
  the critical sets by iteratively removing the Pareto-efficient point
  having minimal Euclidean distance from the ideal point (potential
  ties are broken by selecting one of the points uniformly at
  random). We then recompute the set of Pareto-efficient points and
  iterate until the LMCC becomes non-extensive. Details about the time
  complexity of Pareto-efficient strategies is reported in
  Appendix~\ref{appendix:time_complexity}.}

\section{Comparison of targeted attack strategies}

\newtext{Here we compare the two state-of-the-art algorithms for
  optimal multiplex percolation proposed so far, namely High-Degree
  Adaptive (HDA)~\cite{osat2017optimal} and Effective Multi-Degree
  (EMD)~\cite{Baxter2018targeted}, with a variety of multiplex
  targeted attack strategies from three classes, namely \textit{i)}
  alternative genuinely multiplex strategies; \textit{ii)}
  Pareto-efficient strategies based on the combination of the scores
  of single-layer targeted attack strategies on the two layers; and
  \textit{iii)} Pareto-efficient strategies obtained by combining
  single-layer descriptors with one genuinely multiplex algorithm. 
  In the following we will discuss in detail the performance obtained by
  six strategies, namely HDA, EMD, \DCI, \DCIz, and the two
  Pareto-efficient strategies obtained by combining the k-core ranking
  on the two layers with the ranking induced by EMD, that we call
  (k-core, EMD), and the score assigned on each layer by Believe
  Propagation Decimation and the ranking induced by \DCIz. Notice that
  when considering HDA, we iteratively remove nodes from the duplex that
  have the highest product of the degrees in the two layers, as done in~\cite{osat2017optimal}.  As a
  reference, we also report the results obtained by Simulated
  Annealing (SA) as described in Ref.~\cite{baxter2016correlated},
  which is able to find very small targeted attack sets at the expense
  of relatively heavier computations. The results obtained with all the
  other methods we tested are reported in
  Appendix~\ref{appendix:targeted_strategies}.}

\newtext{In Fig.~\ref{fig:fig1}(a), we report the percolation diagrams
  of duplex networks with uncorrelated \Erdos-\Renyi layers. Notice
  that the duplex network in Fig.~\ref{fig:fig1}(a) is consistent with
  that used in Ref.~\cite{Baxter2018targeted}, where the authors
  showed that the critical set found by EMD is usually much smaller
  than that found using HDA. Interestingly, the combination of EMD and
  k-core provides a smaller critical set than either EMD
  or HDA alone. This is because by targeting nodes which have high EMD
  scores and, at the same time, belong to the inner k-core on each
  layer, we have a higher probability of simultaneously damaging the
  LMCC of the multiplex and the giant connected component on each
  layer. Even more interesting results are reported in
  Fig.~\ref{fig:fig1}(b) for a duplex with maximally disassortative
  inter-layer degree correlations (and no edge overlap), and,
  respectively, in Fig.~\ref{fig:fig1}(c) for a duplex with high edge
  overlap. It is evident from the figures that the relative
  performance of each targeting algorithm depends quite substantially
  on the structure of the multiplex, and in particular on the presence
  of inter-layer degree correlation and edge overlap. For instance,
  EMD still outperforms HDA by a large margin when the multiplex has
  no edge overlap and disassortative degree-degree correlations
  [Fig.~\ref{fig:fig1}(b)], while EMD is the worst-performing strategy
  when edge overlap is not negligible. In general, the algorithms
  based on Pareto-efficiency perform better than either EMD and HDA,
  while both \DCI and \DCIz find relatively smaller critical sets in
  the case of networks with non-negligible overlap. This is a first
  confirmation of our intuition that heuristics that perform better in
  one specific condition (e.g., where the two layers are uncorrelated
  and edge overlap is negligible) do not always achieve the same
  performance under other conditions.}

\subsection{Dependence on edge overlap}
The edge overlap of a two-layer multiplex measures the fraction of
edges that are present on both
layers~\cite{battiston2014,Nicosia2015,Bianconi2018multilayer}. It can
be measured as:
\begin{equation}
o_s = \frac{\sum_{i, j}^{N} o_{i j}}{2 \sum_{i, j}^{N}\Theta\left(o_{i j}\right)},
\end{equation}
where $o_{ij} = \sum_{\alpha=1}^{2} A_{ij}^{[\alpha]}$ and
$\Theta(\bullet)$ is the Heaviside step function. \newtext{In
  particular, $o_s = 1/2$ when the two layers do not share any edge in
  common~\cite{Santoro_complexity2019}. By contrast, the maximum value
  $o_s = 1$ is obtained when the two layers are identical. For the
  sake of convenience, we consider the linear transformation $o=2(o_s
  - 1/2)$ that maps the edge overlap $o_s$ into the interval $[0,
    1]$. In general, real-world multiplex network exhibit relatively
  large values of edge overlap~\cite{battiston2014,Nicosia2015},
  indicating the presence of non-trivial correlations between the two
  layers. Nevertheless, targeted attack strategies have been compared
  mainly (if not exclusively) on duplex networks with \Erdos-\Renyi
  layers having a negligible edge overlap.}

\begin{figure}[!t]
	\includegraphics[width=3in]{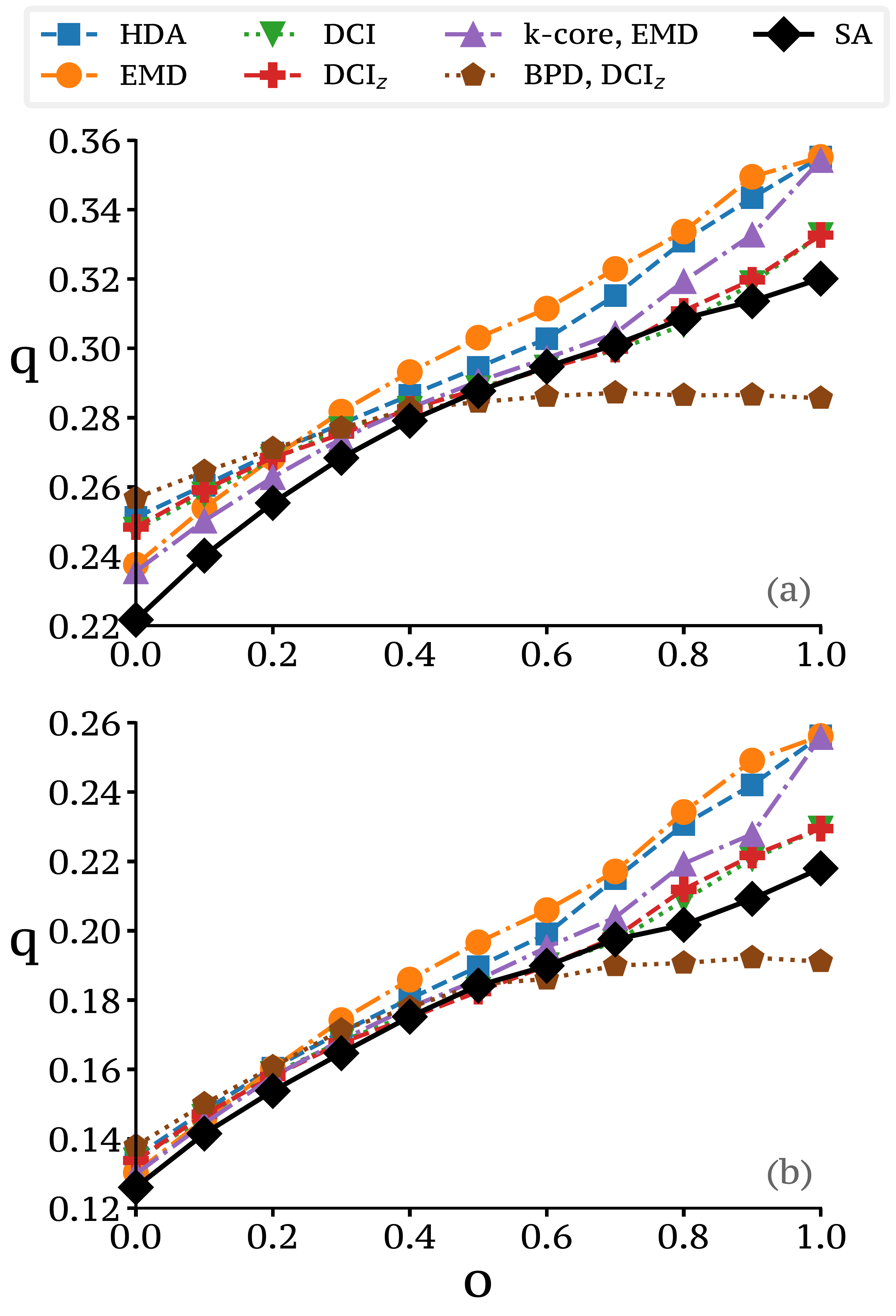}
	\caption{Size of the critical attack set $q$ as a function of edge
    overlap for different attack strategies in duplex networks with
    $N=10^4$ nodes, $\langle k \rangle = 5$, and whose layers have (a)
    \Erdos-\Renyi or (b) scale-free degree distributions
    ($\gamma=2.6$). The plots are obtained by starting from two
    identical layers ($o=1$ and $\rho=1$) and then iteratively
    rewiring the edges of one of the two layers to reduce the edge
    overlap until we get to $o=0$~\cite{Diakonova2016irreducibility}.
    Again, strategies based on Pareto-efficiency yield the best
    results. Results averaged over 20 realisations. Error bars are
    smaller than the marker size.}
	\label{fig:fig2}
\end{figure}

\newtext{We investigated the impact of edge overlap on the performance
  of different targeted attack strategies by considering a class of
  synthetic duplex networks with tunable edge overlap $o$. In
  particular, we employed an approach similar to the one presented
  in~\cite{Diakonova2016irreducibility}. That is, starting from two
  identical layers ($o=1$ and maximal inter-layer degree correlation,
  $\rho =1$), we iteratively rewire the edges of one of the two layers
  to reduce the edge overlap until we get to $o=0$, yet maintaining
  untouched the degree sequence of each layer (see
  Appendix~\ref{appendix:numerical_methods} for details). In
  Fig.~\ref{fig:fig2}(a), we plot the relative size of the critical
  set $q$ obtained by each of the six algorithms as a function of the
  edge overlap in a duplex with \Erdos-\Renyi layers. We notice that in
  general $q$ is an increasing function of $o$. This fact is somehow
  expected, since the existence of an extensive MCC imposes more
  stringent constraints on the graph than the existence of a giant
  connected component in a single-layer graph. Indeed, a duplex with
  $o=1$ is indistinguishable from the single-layer graph obtained by
  combining the two (identical) layers, hence the optimal attack set
  in that case corresponds to that of each layer.}

\newtext{However, each attack strategy behaves slightly differently as
  $o$ increases. For instance, for $o>0.3$ the critical set found by
  EMD is always larger than that obtained by all the other
  strategies. In the limit of $o=1$, however, the EMD and HDA
  heuristics coincide, since the EMD weight of each node $i$ becomes
  proportional to the degree $k_i$. By contrast, \DCI, \DCIz and the
  two Pareto-efficient strategies perform relatively poorly in
  networks with small overlap, but they generally outperform both EMD
  and HDA as the amount of overlap increases. This is because 
  targeted methods that indirectly disrupt a large number of interdependent 
  paths are more likely to trigger cascades in the system.
  Notice that some
  Pareto-efficient strategies already outperform the results of the
  Simulated Annealing achievable in a reasonable computing time (same
  implementation as the one presented in~\cite{Baxter2018targeted}
  with temperature steps equal to $10^{-7}$). A similar qualitative
  behaviour is observed when considering duplex systems having
  heterogeneous degree distribution on each layer
  [Fig.~\ref{fig:fig2}(b)], although the typical values of $q$ are
  overall smaller.  This indicates that the heterogeneity of the
  degree distribution of each layer has some impact on the efficiency
  of each attack strategy, but the presence of edge overlap
  effectively determines the relative performance of different
  strategies. Interestingly, for both the topologies, the best
  (smallest) critical set is always obtained by one of the methods
  proposed in this paper, that is, methods that combine layer-based
  and genuinely multi-layer node properties through Pareto-efficiency,
  with \DCI and \DCIz following closely when $o>0.4$ (see
  Appendix~\ref{appendix:targeted_strategies} for the corresponding
  figure with a comparison of all the targeted attack strategies
  considered).}

\begin{figure*}[!htb]
	\includegraphics[width=0.95\textwidth]{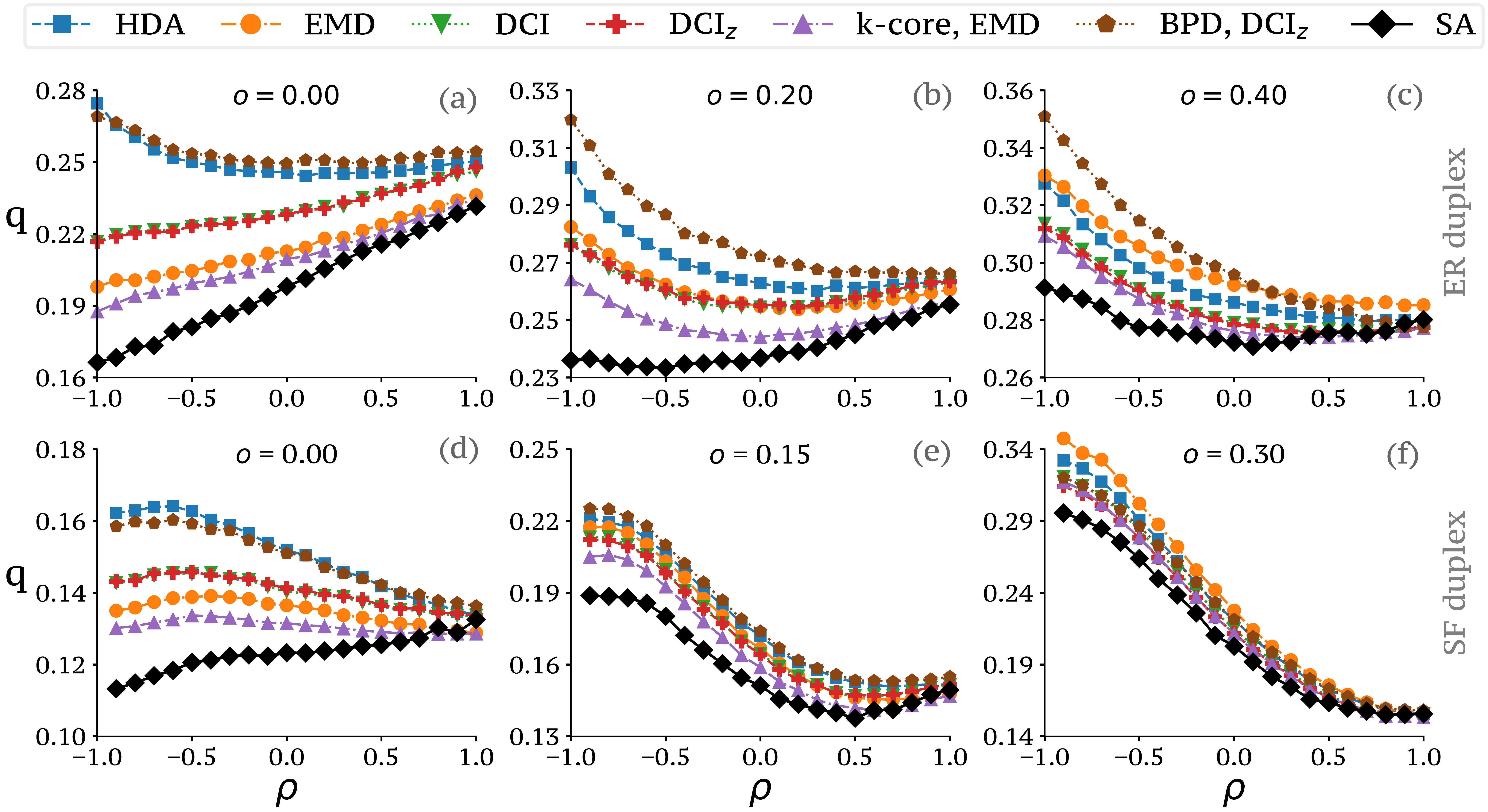}
	\caption{\newtext{Size of the critical set $q$ as a function of the
      inter-layer degree correlations coefficient
      $\rho$~\cite{Nicosia2015} for duplex networks with $N=10^4$
      nodes, whose layers are (a-c) \Erdos-\Renyi graphs with $\langle
      k \rangle = 5$, and (e-f) scale-free networks with $\gamma=2.6$
      (e-f). For each topology, we respectively report three different
      overlap conditions: (a, d) no edge overlap; (b, e) low edge
      overlap; (c, f) moderate edge overlap.} The concurrent presence
    of inter-layer degree correlations and edge overlap strongly
    affects the robustness of a system against targeted
    attacks. \newtext{This is particularly evident when considering
      duplex networks with heterogeneous degree distributions on each
      layer (d-f). When the edge overlap is non-negligible, duplex
      networks with maximally negative degree correlations are
      extremely robust under targeted attacks with respect to their
      maximally positive counterparts. Conversely, when the overlap is
      negligible, disassortatively-correlated duplex networks are more
      fragile. Overall,} \DCI, \DCIz, and the Pareto-efficient
    strategies that simultaneously combine single- and multi-layer
    attacks, consistently detect smaller critical sets. \newtext{The
      results obtained by Simulated Annealing (SA) are reported for
      comparison.} Results averaged over 20 realisations. Error bars
    are smaller than the marker size.}
	\label{fig:fig3}
\end{figure*}

\subsection{The role of inter-layer degree correlations}
Inter-layer degree correlations are known to have a substantial role
in determining the robustness of many real-world
systems~\cite{kleineberg2017geometric}, and are responsible for
consistent shifts in the position of the random percolation
threshold~\cite{Buldyrev_2011_interdependent,Baxter2012avalanche}. Several
studies have found that maximally disassortative inter-layer degree
correlations improve the robustness of multiplex systems to both
random~\cite{Min2014network,kleineberg2017geometric} and targeted
attacks based on the selection of nodes with the largest
degrees~\cite{Min2014network}. However, most of those studies have
only considered the case of duplex systems with identical degree
distributions on the layers, and either maximally positive or
maximally negative inter-layer degree correlations. Here, we used the
procedure explained in Ref.~\cite{Nicosia2015} to tune inter-layer
degree correlations between the maximally disassortative case [also
called Maximally Negative (MN)] and the maximally assortative one
[Maximally Positive (MP)]. In order to isolate the effect of
inter-layer degree correlations, we studied the performance of the
\newtext{six} targeted attack strategies as a function of the
inter-layer degree correlation coefficient $\rho$~\cite{Nicosia2015},
imposing that each realisation of the multiplex had $o\approx 0$.  To
simultaneously account for the joint effect of overlap and inter-layer
degree correlations, we also consider the sequences of multiplex
networks obtained by increasing $\rho$ while keeping the edge overlap
fixed at a given value. To obtain those sequences, we first increase
the value of inter-layer degree correlation~\cite{Nicosia2015}, and
then we set the desired value of edge overlap through biased edge
rewiring~\cite{Diakonova2016irreducibility,Santoro_complexity2019}
(see Appendix~\ref{appendix:numerical_methods} for a more detailed
description of the method).

\newtext{In Fig.~\ref{fig:fig3} we plot the size of the critical set
  $q$ identified by the six targeted attack strategies as a function
  of the inter-layer degree correlations $\rho$ and for different
  values of edge overlap $o$. We report the results obtained on duplex
  networks with \Erdos-\Renyi layers [Fig.~\ref{fig:fig3}(a-c)], and
  with scale--free layers [Fig.~\ref{fig:fig3}(d-f)].  Interestingly,
  in all the scenarios considered the state-of-the-art EMD and HDA are
  outperformed by one or more of the heuristics proposed in this
  paper. In particular, the smallest critical set is often obtained by
  the (k-core, EMD) Pareto-efficient strategy. However, depending on
  the interplay between edge overlap and inter-layer degree
  correlations, profound differences among the six methods emerge.
  For instance, when considering a duplex with \Erdos-\Renyi layers and
  negligible edge overlap [Fig.~\ref{fig:fig3}(a)], the discrepancy
  between the overall best strategy (k-core, EMD) and the worst one
  (HDA) is maximal when $\rho\simeq -1$}. In particular, the critical
set found by (k-core, EMD) when $\rho\simeq -1$ is around $19\%$
(smaller than the one found for $\rho\simeq 1$, i.e., around $23\%$),
while HDA finds a much larger critical set ($28\%$), which is even
larger than the one it finds for $\rho\simeq 1$ ($25\%$). \newtext{As
  a consequence, the presumed increased robustness of multiplex
  networks with disassortatively correlated degrees is probably just
  an artefact of the algorithm used to determine the critical
  set~\cite{Min2014network}. By looking at the size of the critical
  set found by Simulated Annealing in Fig.~\ref{fig:fig3}(a), it seems
  clear that negatively-correlated multiplex systems without overlap
  are generally hyperfragile compared to positively-correlated
  ones. However, some of the attack strategies considered, including
  HDA and especially in uncorrelated systems, provide a diametrically
  opposite picture, and suggest that in absence of edge overlap
  positively correlated degree sequences are more fragile. The results
  shown in Fig.~\ref{fig:fig3}(b)-(c) shed light on the interplay
  between edge overlap and inter-layer correlations. In both cases,
  the sizes of the critical sets found by the six algorithms are
  higher than those shown in Fig.~\ref{fig:fig3}(a) (i.e., when the
  edge overlap is negligible). In particular, the (sub-)optimal
  critical set $q$ found by Simulated Annealing reveals that both edge
  overlap and inter-layer degree correlations contribute to determine
  the robustness of a duplex system.}

\begin{figure}[t!]
	\includegraphics[width=0.48\textwidth]{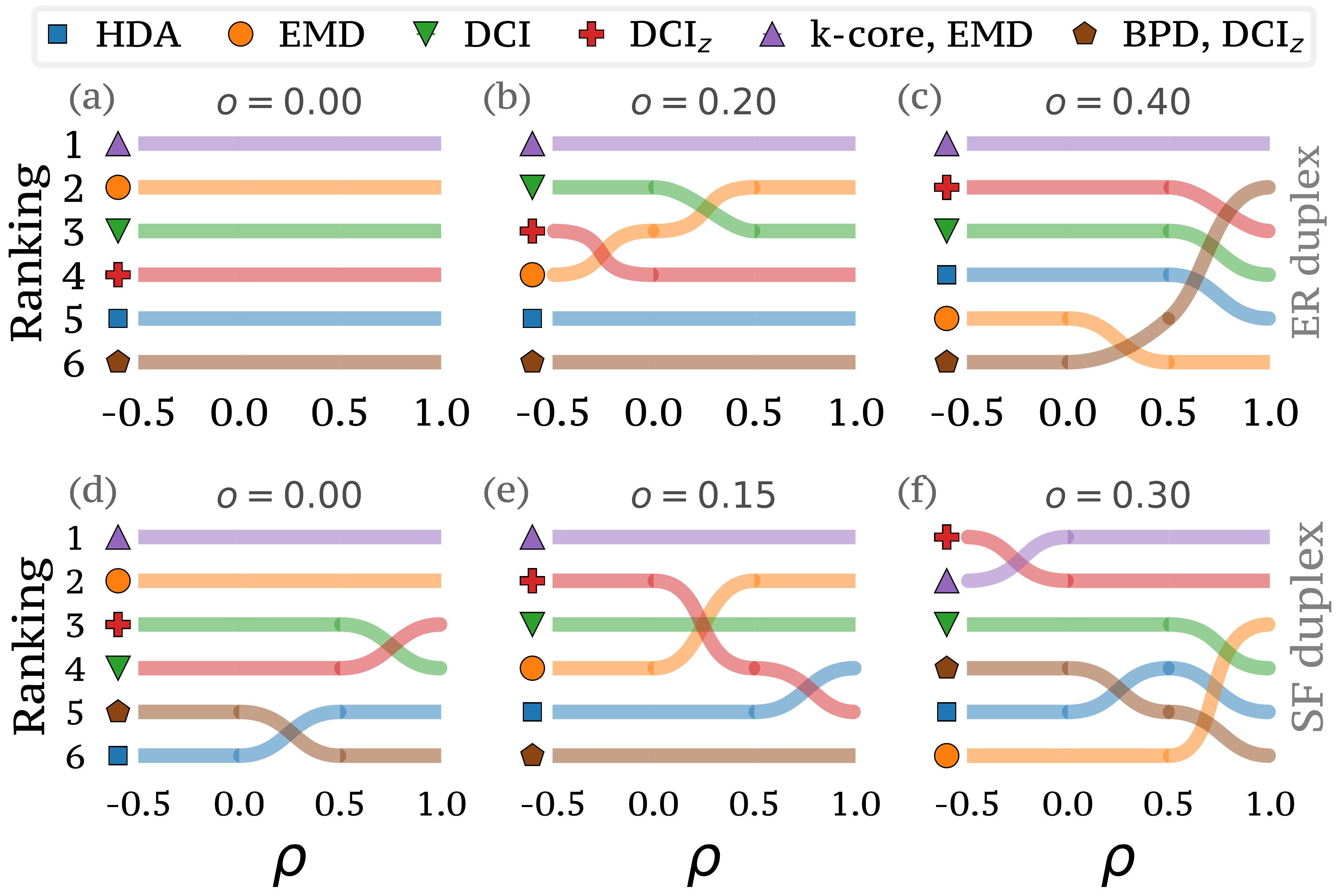}
	\caption{\newtext{Rankings of the six targeted attack strategies
      presented in Fig.~\ref{fig:fig3} for different values of edge
      overlap and different network topology, namely, \Erdos-\Renyi
      (a-c) and scale--free (d-f), as a function of four values of
      inter-layer degree correlations. Interestingly, the (k-core, EMD)
      Pareto-efficient strategy has a considerably better performance
      in most of the conditions considered. By contrast, the EMD
      strategy appears to have a good performance only in duplex
      networks with negligible or small overlap.}}
	\label{fig:fig4}
\end{figure}

\newtext{Similar conclusions can be drawn by examining duplex networks
  with scale--free degree distribution [see
    Fig.~\ref{fig:fig3}(d-f)]. Also in this case both edge overlap and
  inter-layer degree correlations have a substantial impact on the
  performance of each algorithm. However, the difference between
  maximally-negative and maximally-positive correlated duplex networks
  becomes more relevant when edge overlap increases, mainly due to the
  fact that degree heterogeneity on each layer has a stronger impact
  on the percolation of the MCC. It is interesting to notice here
  that, since the relative performance of the algorithms considered
  clearly depends on both inter-layer degree correlations and edge
  overlap, there is no algorithm that clearly outperforms all the
  others. This is made evident in Fig.~\ref{fig:fig4}, where we
  highlight the behaviour of the ranking of the six heuristics based
  on increasing size of the critical set $q$ (i.e., the algorithm
  ranked first is the one providing the smallest critical
  set). Although the (k-core, EMD) Pareto-efficient strategy seems to
  perform consistently well across the board, being ranked first or
  second more often that the other five strategies, there are several
  combinations of layer structure, edge overlap, and inter-layer
  degree correlations for which other algorithms perform much better.
  An overview of the critical sets found by all the attack strategies
  we have considered as a function of overlap and correlations is
  reported in Appendix~\ref{appendix:targeted_strategies}.}

\section{Optimal percolation in real-world multiplex networks}

\begin{figure}[t]
	\includegraphics[width=0.49\textwidth]{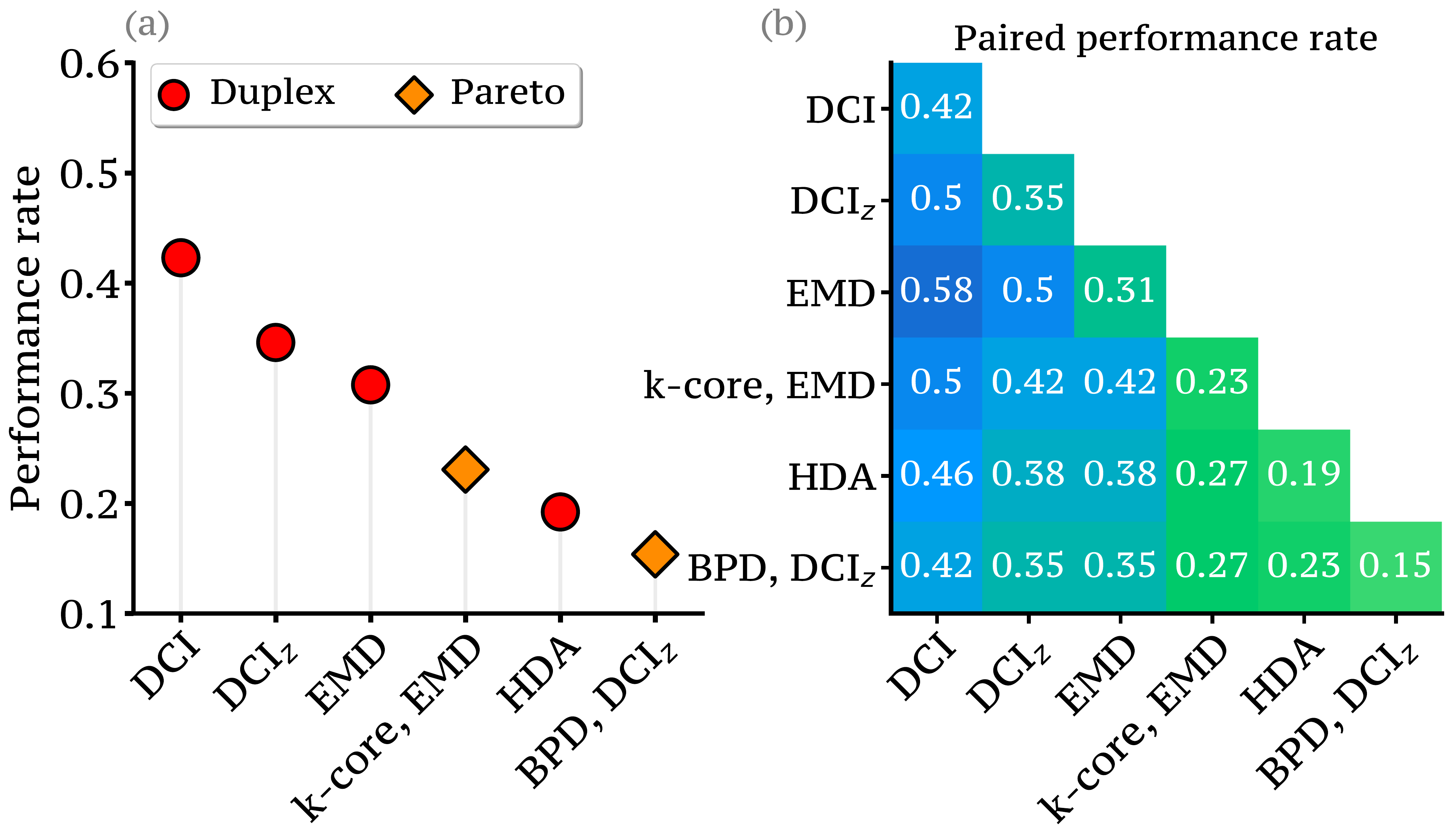}
	\caption{\newtext{(a) Relative performance of the six targeted
      attack strategies for real-world duplex networks. Notice that
      strategies that perform well in synthetic duplex networks are
      not the best ones when it comes to real-world systems. In panel
      (b) we show the overall pair performance rate, defined as the
      relative number of times either of a pair of algorithms
      identifies the smallest critical set. In this case, the
      combination of \DCI and \DCIz with other strategies results in
      the best performance rates. However, the fact that the highest
      value of pair performance is 58\% suggests that we are still far
      from having fully understood the robustness of real-world
      systems to targeted attacks. Results are averaged over 10
      realisations.}}
	\label{fig:fig5}
\end{figure}

\newtext{One of the main aims behind the study of targeted attacks is
  to try to find efficient ways to mitigate the fragility of
  real-world infrastructures, which are normally characterised by
  layer heterogeneity, non-negligible edge overlap, and inter-layer
  degree correlations. For this reason, we tested the targeted attack
  strategies presented in this paper in 26 real-world multiplex
  networks~\cite{structural_reducibility_2015}. The size of the
  critical set found by each of the algorithms is reported in
  Table~\ref{table:real_multiplex}. The systems considered in the
  Table range in size from few dozens to thousands of nodes, with
  different values of edge overlap and inter-layer degree
  correlations. Since many of those multiplex networks have more than
  two layers, for each system we considered the duplex sub-networks
  corresponding to the pairs of layers yielding the largest MCC, as
  already done for instance in the main text of Ref.~\cite{kleineberg2017geometric}. }
Unsurprisingly, there is no single strategy that works better than all
the others in all the cases. \newtext{What is
  surprising instead is that those strategies yielding the best
  performance when considering synthetic duplex systems, e.g., the
  (k-core, EMD) Pareto-efficient algorithm, do not perform as well in
  real-world systems. By contrast, \DCI and \DCIz quite often find
  the smallest critical set. This can be
  easily visualised in Fig.~\ref{fig:fig5}(a), where we plot the
  relative amount of times (i.e. performance rate) that a certain
  strategy identifies the smallest critical set in all the 26
  real-world duplex networks considered. The best-performing strategy
  here is \DCI, with a rate of 42\%, followed by \DCIz (35\%) and EMD
  (31\%). We also considered the pair performance, that is defined as
  the relative number of times that at least one of two algorithms
  identifies the smallest critical set.  The results are reported in
  Fig.~\ref{fig:fig5}(b). Remarkably, combinations of targeted attacks
  including \DCI and \DCIz yield the best pair performance rate, where
  the pair [\DCI, EMD] is able to find the smallest critical set in
  58\% of the cases. Overall, these results warn against the quest to
  find a single targeted attack strategy that performs well whatever
  the multiplex network it is applied to. In particular, the
  generalisation to real-world networks of targeting strategies that
  perform well in specific classes of synthetic graphs can result in
  the gross overestimation of the robustness of a system.}
\section{Conclusions}
Optimal multiplex percolation is characterised by a variety of
subtleties, and we are probably far from having already understood it
completely. The massive comparison of many different attack strategies
that we presented here has revealed that the performance of all the
state-of-the-art attack strategies on uncorrelated multiplex networks
does not guarantee their ability to identify sufficiently small
critical sets in networks having non-negligible edge overlap and
non-trivial inter-layer degree correlations. \newtext{In particular,
  extensive simulations on synthetic networks have shown that both
  edge overlap and inter-layer degree correlations play an important
  role in determining the robustness of a system, and that their
  combination can be effectively exploited to modulate the robustness
  of a system against targeted attacks. One of the main ingredients to
  identify critical nodes using only local information seems to be
  their potential to disrupt a lot of paths among their
  second-neighbours, since this would indirectly contribute to the
  disruption of the LMCC. The Duplex Collective Influence algorithm
  presented here is based on this assumption, and indeed shows a
  relatively better performance than all the existing state-of-the-art
  algorithms, especially in duplex networks with non-negligible edge
  overlap. It seems clear to us that further improvements might
  possibly be obtained by considering extensions of \DCI that take
  into account the impact on farther-away neighbourhoods. Another
  important ingredient seems to be the possibility to combine
  structural descriptors on each of the two layers with more
  genuinely-multiplex information. In this respect, the family of
  Pareto-efficient strategies that we presented here represents a
  quite promising approach. We find it remarkable that these
  strategies consistently outperform all the existing algorithms in
  both synthetic and real-world duplex networks.}

\newtext{One of the main motivations behind studying percolation is to
  improve our ability to assess the robustness and to mitigate the
  fragility of real-world networks, i.e., of concrete systems
  presenting non-negligible edge overlap and non-trivial inter-layer
  degree correlations. And the most surprising results indeed came
  from the analysis of real-world multiplex systems, and provide a
  clear warning against hasty generalisations. On the one hand, the
  heuristics that are good at finding small critical sets in
  uncorrelated multiplex networks often perform rather poorly in
  real-world systems, thus tending to overestimate their
  robustness. On the other hand, the variability in performance shown
  by almost all the algorithms we have considered confirms that a fair
  assessment of the robustness of a multi-layer system must be based
  on the usage and comparison of multiple attack strategies. We
  believe that these results constitute a solid spring-board for a
  more in-depth investigation of optimal percolation in multi-layer
  systems.}

\section*{Code Availability}
Implementations of the 20 targeted attack strategies used in this paper, and
of the algorithm to tune overlap and inter-layer degree correlations in
synthetic duplex systems, are available at~\cite{Santoro_percolation}:
\url{https://github.com/andresantoro/Multiplex_optimal_percolation}

\section*{Acknowledgments}
V. N. acknowledges support from the EPSRC Grant
No. EP/S027920/1. A. S. acknowledges support from The Alan Turing
Institute under the EPSRC Grant No. EP/ N510129/1.  This work made use
of the MidPLUS cluster, EPSRC Grant No. EP/K000128/1.

\begin{table*}[h]
{\renewcommand{\arraystretch}{1.6}
		\scriptsize
	\noindent\(\begin{tabular}{llcclllllllllllllllllll}
 \text{\textbf{Data set}}  & \text{\textbf{MCC}} & \text{$\mathbf{\rho}$} & \text{$\textbf{o}_{norm}$} & \rot{HDA} & \rot{EMD} & \rot{\DCI} & \rot{ \DCIz}
& \rot{k-core, EMD} & \rot{k-core, \DCI} & \rot{k-core, \DCIz} & \rot{CoreHD, EMD} & \rot{CoreHD, \DCI} & \rot{CoreHD, \DCIz}
& \rot{CI$_{\ell =2}$, EMD} & \rot{CI$_{\ell =2}$, \DCI} & \rot{CI$_{\ell =2}$, \DCIz} & \rot{BPD, EMD} & \rot{BPD, \DCI} & \rot{BPD, \DCIz} & \rot{CoreHD, CoreHD} & \rot{CI$_{\ell =2}$, CI$_{\ell =2}$} & \rot{BPD, BPD} \\
\hline \hline \\[-0.5 em]
\text{Air. FR-U2} & 28 & 0.25 & 0.14 & \underline{4} & 5 &  \underline{4} & \underline{4} & \underline{4} & \underline{4} & \underline{4} & \underline{4} & \underline{4} & \underline{4} & \underline{4} & \underline{4} & \underline{4} & 5 &  \underline{4} & 5 &  \underline{4} & \underline{4} & 5\\
\text{Air. AA-DL} & 191 & 0.72 & 0.60 & \underline{8} & \underline{8} & \underline{8} & \underline{8} & \underline{8} & \underline{8} & \underline{8} & \underline{8} & \underline{8} & \underline{8} & \underline{8} & \underline{8} & \underline{8} & \underline{8} & \underline{8} & \underline{8} & 9 &  \underline{8} & 9\\
\text{Air. AA-UA} &  204 & 0.77 & 0.58 & 9 &  \underline{7} & \underline{7} & \underline{7} & 8 &  \underline{7} & \underline{7} & 8 &  8 &  8 &  9 &  8 &  9 &  9 &  8 &  8 &  9 &  9 &  8\\
\text{UK Train L 26-41} & 59 & 0.38 & 0.19 &  \underline{5} & \underline{5} & \underline{5} & \underline{5} & \underline{5} & \underline{5} & \underline{5} & \underline{5} & \underline{5} & \underline{5} & \underline{5} & \underline{5} & \underline{5} & \underline{5} & \underline{5} & \underline{5} & \underline{5} & \underline{5} & \underline{5}\\
\text{UK Train L 30-41} & 43 & 0.19 & 0.18 & \underline{1} & \underline{1} & \underline{1} & \underline{1} & \underline{1} & \underline{1} & \underline{1} & \underline{1} & \underline{1} & \underline{1} & \underline{1} & \underline{1} & \underline{1} & \underline{1} & \underline{1} & \underline{1} & \underline{1} & \underline{1} & \underline{1}\\
\text{ArXiv L 2-6} & 916 & 0.85 & 0.77 & \underline{91} & 119 &  106 &  106 &  139 &  108 &  108 &  115 &  101 &  100 &  101 &  104 &  103 &  137 &  121 &  115 &  117 &  104 &  119\\
\text{ArXiv L 3-6} & 790 & 0.92 & 0.81 & 85 &  \underline{84} & 98 &  98 &  115 &  103 &  103 &  96 &  99 &  99 &  85 &  97 &  97 &  106 &  100 &  96 &  106 &  87 &  107\\
\text{CS Aarhus L 1-5} & 58 & 0.32 & 0.63 & 14 &  15 &  15 &  \underline{12} & 13 &  15 &  15 &  \underline{12} & 15 &  15 &  \underline{12} & 14 &  14 &  15 &  15 &  13 &  18 &  14 &  14\\
\text{FAO L 3-24} & 193 & 0.94 & 0.70 &83 &  85 &  87 &  85 &  86 &  90 &  90 &  \underline{82} & 88 &  88 &  \underline{82} & \underline{82} & \underline{82} & \underline{82} & 86 &  88 &  88 &  \underline{82} & 85\\
\text{IMDb Com.-Dra.} & 181 & 0.97 & 0.82 & 88 &  90 &  \underline{87} & 89 &  88 &  90 &  90 &  91 &  90 &  90 &  88 &  90 &  89 &  90 &  89 &  89 &  89 &  88 &  89\\
\text{Terr. L Tru.-Op.} &  61 & 0.22 & 0.50 & 18 &  16 &  \underline{15} & \underline{15} & 19 &  19 &  19 &  17 &  17 &  17 &  19 &  19 &  19 &  \underline{15} & \underline{15} & \underline{15} & 17 &  21 &  21\\
\text{Arabid. L 1-2} & 442 & 0.65 & 0.40 &34 &  32 &  29 &  30 &  34 &  31 &  31 &  31 &  31 &  31 &  27 &  31 &  31 &  35 &  39 &  34 &  33 &  35 &  \underline{21}\\
\text{Drosoph. L 1-2} & 299 & 0.18 & 0.07 & 10 &  9 &  14 &  12 &  10 &  14 &  14 &  11 &  \underline{8} & 12 &  9 &  \underline{8} & 10 &  \underline{8} & 10 &  10 &  10 &  12 &  9\\
\text{Drosoph. L 1-3} & 202 & 0.27 & 0.06 & 4 &  4 &  \underline{3} & \underline{3} & 5 &  4 &  4 &  \underline{3} & \underline{3} & \underline{3} & \underline{3} & \underline{3} & \underline{3} & \underline{3} & 4 &  4 &  \underline{3} & \underline{3} & \underline{3}\\
\text{Drosoph. L 1-4} & 1024 & 0.13 & 0.09 & 26 &  26 &  27 &  26 &  35 &  39 &  39 &  25 &  37 &  31 &  \underline{24} & 27 &  30 &  38 &  42 &  38 &  40 &  33 &  42\\
\text{Drosoph. L 2-3} & 449 & 0.68 & 0.35 & 49 &  48 &  50 &  53 &  52 &  50 &  50 &  \underline{47} & 48 &  49 &  55 &  51 &  51 &  53 &  54 &  51 &  54 &  54 &  52\\
\text{Homo L 1-2} & 9312 & 0.57 & 0.32 &1052 &  1058 &  1026 &  1021 &  \underline{1017} & 1045 &  1045 &  1042 &  1051 &  1069 &  1088 &  1060 &  1095 &  1088 &  1060 &  1092 &  1087 &  1101 &  1172\\
\text{Homo L 1-5} & 3886 & 0.31 & 0.16 & 312 &  299 &  \underline{271} & 297 &  301 &  315 &  315 &  307 &  305 &  319 &  344 &  332 &  347 &  329 &  347 &  358 &  341 &  363 &  432\\
\text{Homo L 2-5} & 4944 & 0.44 & 0.17 & 420 &  \underline{366} & 406 &  387 &  389 &  413 &  413 &  386 &  410 &  395 &  412 &  434 &  422 &  410 &  439 &  435 &  481 &  473 &  534\\
\text{Hum. HIV L 1-2} & 144 & 0.54 & 0.41 &  4 &  4 &  3 &  3 &  4 &  3 &  3 &  4 &  \underline{2} & \underline{2} & 4 &  4 &  4 &  4 &  3 &  3 &  \underline{2} & 4 &  3\\
\text{Mus L 1-3} & 1059 & 0.56 & 0.37 & 60 &  60 &  52 &  \underline{50} & 57 &  52 &  52 &  61 &  69 &  65 &  58 &  57 &  60 &  61 &  68 &  56 &  60 &  58 &  63\\
\text{S. Cerev. L 1-2} & 4531 & 0.36 & 0.10 & 785 &  757 &  768 &  769 &  \underline{743} & 785 &  785 &  795 &  782 &  787 &  853 &  823 &  831 &  811 &  811 &  825 &  859 &  885 &  956\\
\text{S. Cerev. L 1-7} & 4720 & 0.28 & 0.07 & 982 &  940 &  \underline{898} & 905 &  911 &  930 &  930 &  994 &  958 &  942 &  1108 &  1000 &  1047 &  1025 &  953 &  973 &  1009 &  1025 &  1115\\
\text{S. Pombe L 3-4} & 1112 & 0.20 & 0.14 & 56 &  \underline{41} & 50 &  52 &  54 &  59 &  59 &  44 &  54 &  52 &  54 &  54 &  53 &  57 &  62 &  60 &  72 &  79 &  75\\
\text{S. Pombe L 3-6} & 956 & 0.14 & 0.06 & 39 &  \underline{32} & 40 &  35 &  42 &  38 &  38 &  35 &  40 &  35 &  38 &  38 &  37 &  45 &  41 &  39 &  50 &  49 &  58\\
\text{S. Pombe L 4-6} & 2292 & 0.61 & 0.01 & 370 &  361 &  \underline{353} & 360 &  358 &  366 &  366 &  377 &  369 &  366 &  385 &  371 &  376 &  383 &  377 &  372 &  382 &  380 &  404\\\hline \hline
	\end{tabular}\)
}
\caption{\newtext{Size of the attack sets for all the different
    strategies considered in this paper for 26 different real-world
    duplex
    networks~\cite{structural_reducibility_2015,Santoro_complexity2019}. For
    each data set we report the size of the initial MCC, the value of
    inter-layer degree correlations within the MCC ($\rho$), and the
    normalised edge overlap ($o_{\rm~norm}=o/o_{\rm max}$ where
    $o_{\max}$ is the maximum overlap in the corresponding
    configuration model
    ensemble~\cite{Diakonova2016irreducibility,Santoro_complexity2019}
    (see Appendix~\ref{appendix:numerical_methods} for details). The
    performance of each strategy heavily depends on the presence of
    edge overlap and inter-layer degree correlations. Interestingly,
    \DCI, \DCIz, and Pareto-efficient strategies based on them perform
    better than all the other strategies. For methods that rely on
    random tie-breaking, the numbers reported correspond to the
    minimum value found by the method over 10 independent
    realisations. The size of the minimal attack set found on each
    duplex is underlined.}}
\label{table:real_multiplex} 
\end{table*}

\appendix

\section{Additional details on \DCI and \DCIz}
\label{appendix:DCI}
\noindent
\newtext{\textbf{Dependence of Duplex Collective Influence score on
    edge overlap. -- } Here we study the character of the \DCI score
  of a node $i$ as a function of $k_i^{\rm int}$, that is the degree
  of node $i$ in the intersection graph, obtained by considering all
  and only the links that exist on both layers. Notice that $k_i^{\rm
    int}$ is intimately connected to the edge overlap around node
  $i$. Indeed, the fraction of edges attached to node $i$ that exist
  in both layers can be expressed as $o_i = k_i^{\rm int}/k_i^{\rm
    aggr}$.  Since we have $k_i^{\rm aggr} = k_i\lay{1} + k_i\lay{2} -
  k_i^{\rm int}$, the \DCI score of node $i$ can be rewritten as:
\begin{equation*}
  DCI(i)\!=\!\! \frac{k_i\lay{1} k_i\lay{2}-k_i^{\rm int}}{k_i\lay{1} +
    k_i\lay{2} - k_i^{\rm int}}\!\!\left[\!\sum_{j} a_{ij}\lay{1}
    (k_j\lay{2} - 1) + a_{ij}\lay{2}(k_j\lay{1} - 1)\!\right]
\end{equation*}
The term inside square brackets does not depend on $k_i^{\rm int}$, so
that we can just focus on the ratio outside, which can be
conveniently rewritten as:
\begin{equation}
  \frac{a - k_i^{\rm int}}{b - k_i^{\rm int}}
  \label{eq:ratio}
\end{equation}
where we have set $a=k_i\lay{1}k_i\lay{2}$ and
$b=k_i\lay{1}+k_i\lay{2}$.  Notice that $k_i^{\rm int}\in \left[0,
  \min\left(k_i\lay{1}, k_i\lay{2}\right)\right]$ and in particular
$k_i^{\rm int}=0$ if the neighbourhoods of node $i$ at the two layers
are disjoint, while $k_i^{\rm int} = \min \left(k_i\lay{1},
k_i\lay{2}\right)$ if the intersection between those two
neighbourhoods is maximal, where the case $k_i\lay{1} = k_i\lay{2}$
corresponds to identical neighbourhoods on the two layers. It is easy
to show that Eq.~(\ref{eq:ratio}) is an increasing function of
$k_i^{\rm int}$ for $a > b$, which holds whenever
$\min\left(k_i\lay{1}, k_i\lay{2}\right)>1$. This means that, all
other things being equal, a node having degree larger than one on both
layers will have a larger \DCI score if it has a larger edge
overlap. A similar reasoning holds for \DCIz.}

\medskip
\noindent
\newtext{\textbf{DCI in multiplex networks with identical layers. --}
  It is easy to show that in a duplex network with identical layers
  the ranking of nodes induced by the \DCI score defined in
  Eq~(\ref{eq:DCI_formula}) coincides with that induced by the CI
  score on the corresponding aggregated graph when $\ell=1$. In fact, if the two
  layers are identical, we have $a_{ij}\lay{1} = a_{ij}\lay{2} =
  a_{ij} \> \forall i, j=1,\ldots, N$ and also $k_i\lay{1} =
  k_i\lay{2} = k_i^{\rm int} = k_i^{\rm aggr} = k_i \> \forall
  i=1,\ldots,N$, so we get $DCI(i) = 2(k_i -1) \sum_{j} a_{ij}(k_j -
  1) = 2 CI_{\ell=1}(i)$, which means that the two rankings are identical.}

\medskip
\section{Time complexity}
\label{appendix:time_complexity}

\noindent
\newtext{\textbf{Time complexity of \DCI and \DCIz. --} The adaptive
  targeted strategies based on \DCI and \DCIz require to re-compute
  the \DCI scores of all the remaining nodes after each node is
  removed. An implementation with simple data structures (basically,
  the list of neighbours of each node) guarantees a worst-case time
  complexity $\mathcal{O}(N^2 \log N)$, where $N$ is the number of nodes of the
  graph. Indeed, the initial \DCI (or \DCIz) score of all the nodes
  can be computed in $\mathcal{O}(K)$ (where $K$ is the total number of edges of
  the multiplex), and sorted in $\mathcal{O}(N \log N)$. The removal of the
  $i$-th node from the network will modify the \DCI scores of all its
  neighbours on the two layers, which are at most $N-i-1$. Since we
  need to keep the list of \DCI scores ordered, the usage of simple
  structures requires to sort again the scores, which has time
  complexity $\mathcal{O}(N \log N)$ at each step. As a consequence, updating
  \DCI scores throughout the percolation procedure has time complexity
  $\mathcal{O}(N^2 \log N)$. A direct computation and update of the size of the
  LMCC would run in $(N^3)$, but its efficiency can be improved to
  $\mathcal{O}(N^{1.2})$ by using the algorithm explained in Ref.~\cite{Hwang2015efficient_MCC, Lee2018decremental}. So overall the \DCI (\DCIz) algorithm for \DCI and
  \DCIz has time complexity $\mathcal{O}(N^2 \log N)$. However, the usage of a
  max-heap to store and update the list of \DCI scores would guarantee
  a worst-case time complexity of $\mathcal{O}(N^{1.2} + K \log N)$, which is
  dominated by $\mathcal{O}(N^{1.2})$ in sparse graphs.}

\begin{figure*}[!htb]
\includegraphics[width=0.95\textwidth]{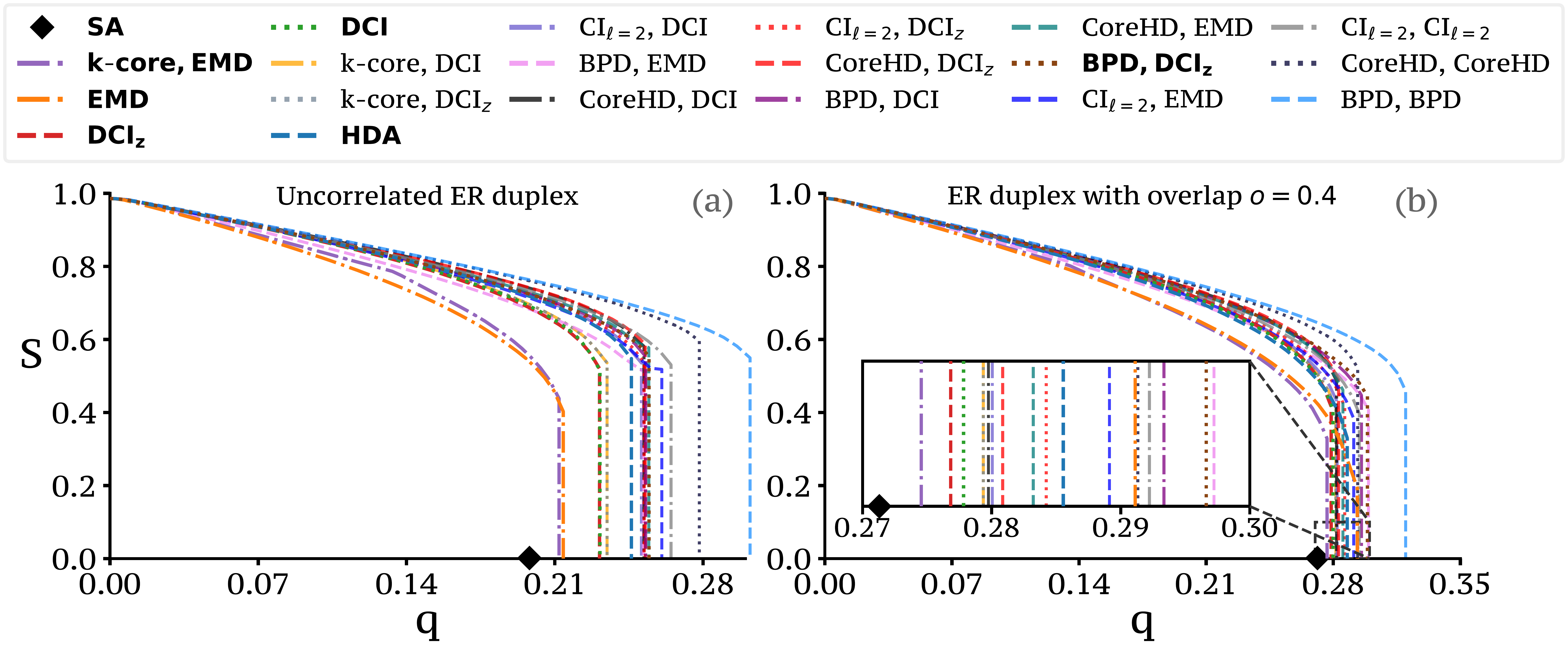}
\caption{\newtext{Percolation diagram for the 20 different attack
    strategies examined in this paper, for the same duplex networks
    respectively reported in Fig.~\ref{fig:fig1}(a) and in
    Fig.~\ref{fig:fig1}(c). Notice that Pareto-efficient strategies
    combining only single-layer metrics do not perform well in the two
    cases reported. By contrast, methods which effectively combine
    single- and multi-layer information yield the best
    performance. Labels are sorted in ascending order of size of the
    critical set from panel (a). We highlighted in bold the six
    strategies presented in Fig.~\ref{fig:fig1}. Results averaged over
    20 realisations.}}
\label{fig:percolation_diagram_all}
\end{figure*}
\begin{figure}[!htbp]
\includegraphics[width=0.48\textwidth]{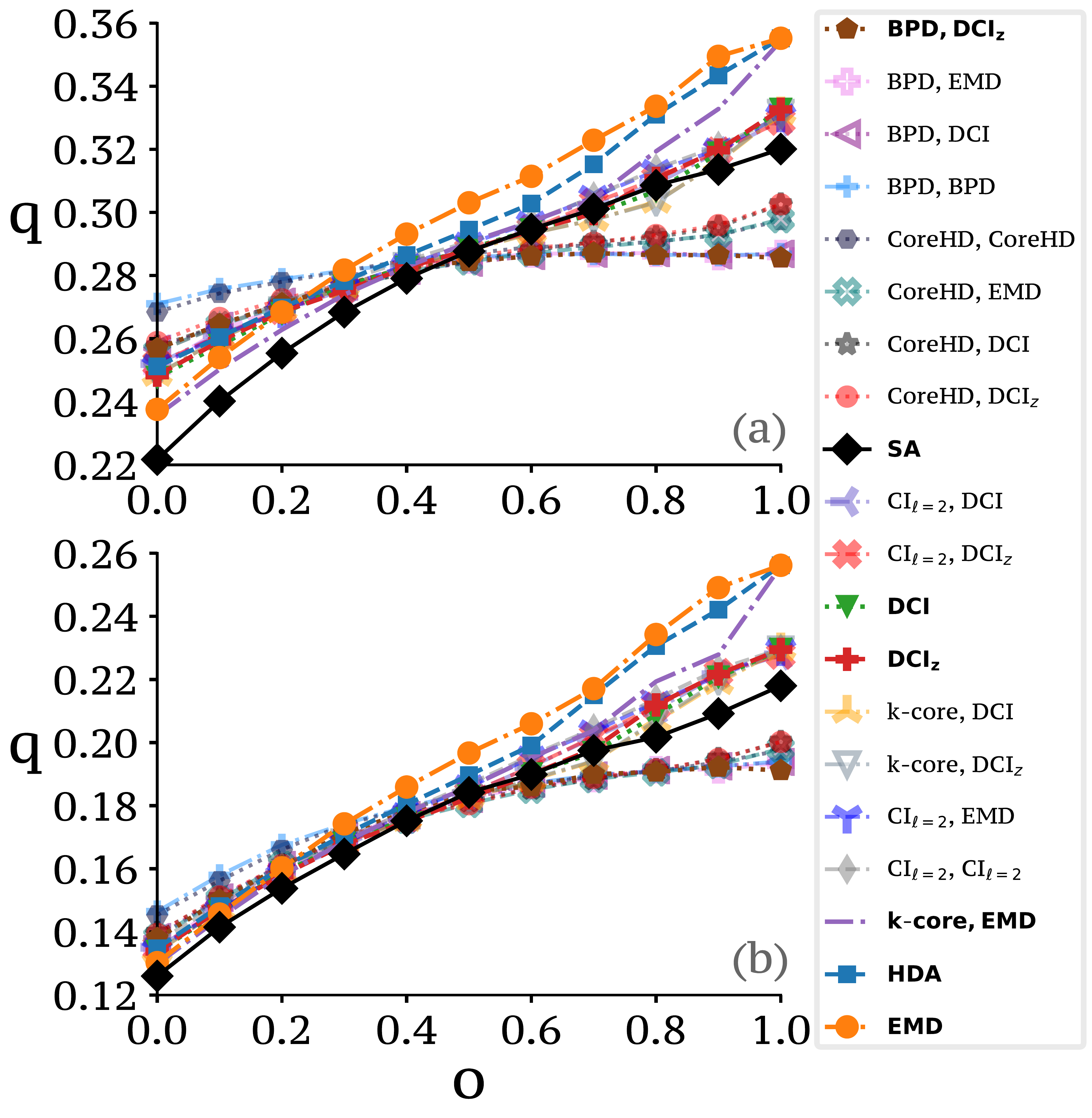}
\caption{\newtext{Size of the critical attack set $q$ as a function of
    edge overlap for the 20 different attack strategies in the same
    duplex networks as in Fig.~\ref{fig:fig2}. Labels are sorted  in
    ascending order of size of the critical set from panel (a) when
    $o=1$ [i.e., it is analogous to the single-layer percolation].
     Results averaged over 20 realisations. Error bars are smaller than the marker
    size.}}
\label{fig:percolation_vs_overlap_all}
\end{figure}

\medskip
\noindent
\newtext{\textbf{Time complexity of Pareto-efficient strategies. -- }
  The time complexity of Pareto-efficient strategies can be expressed
  as $\mathcal{O}(F + S)$, where $\mathcal{O}(S)$ is the time complexity of computing and
  updating the scores used for multi-objective optimisation, while
  $\mathcal{O}(F)$ is the time complexity of computing and updating the
  Pareto-front throughout the percolation procedure. Identifying the
  Pareto Front at each step has time complexity $\mathcal{O}(N \log
  N)$ when the number of objective functions $m$ is at most $m=3$,
  which is the case for all the Pareto-efficient strategies considered
  in the present paper. If the number of functions to optimise is
  $m>3$, then the time complexity becomes $\mathcal{O}(N (\log
  N)^{m-2})$~\cite{Kung_75}. As a consequence $\mathcal{O}(F) = \mathcal{O}(N^2 \log N)$
  in the worst case. The time complexity of computing and updating the
  scores depends on the details of the functions used, but all the
  functions we used in this paper are dominated by $\mathcal{O}(N^2 \log N)$.}

\begin{figure*}[!htbp]
    \includegraphics[width=0.95\textwidth]{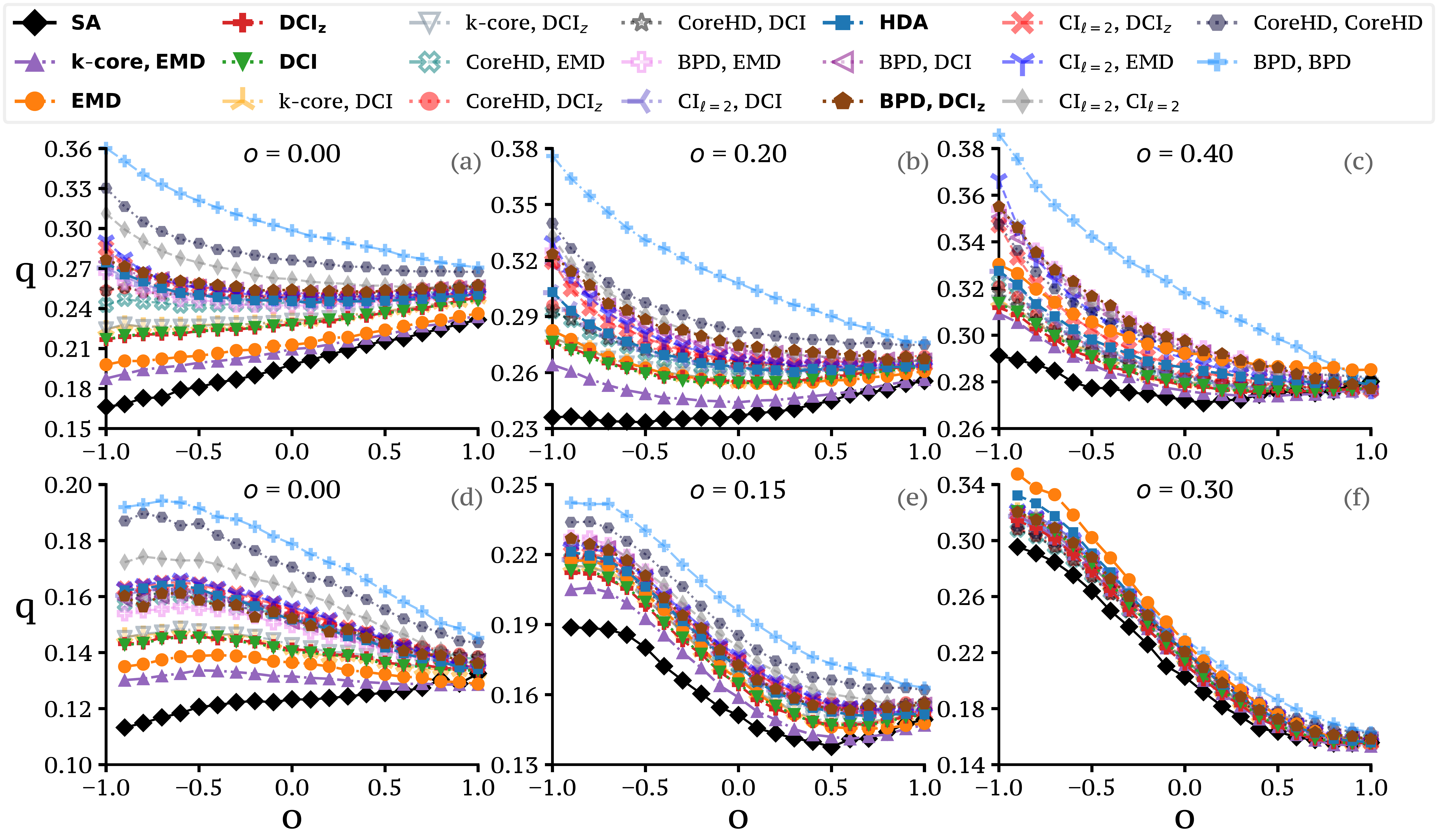}
\caption{\newtext{Size of the critical set $q$ found by each of the 20
    attack strategies as a function of the inter-layer degree
    correlations coefficient for the same duplex networks as in
    Fig.~\ref{fig:fig3}. Notice that also in this case the attack
    strategies which combine only single-layer metrics perform quite
    poorly compared to the other Pareto-efficient methods. Labels are
    sorted in ascending order of $q$ when considering $\rho=-1$ of
    panel (a), while in bold we highlighted the methods presented in
    Fig.~\ref{fig:fig3}. Results averaged over 20 realisations. Error
    bars are smaller than the marker size.}}
\label{fig:percolation_vs_rhocorrelation_all}
\end{figure*}

\section{Additional results on synthetic networks}
\label{appendix:targeted_strategies}

\newtext{In this section we report the results obtained by the
  multiplex targeted strategies constructed by considering all
  Pareto-efficient combinations of different of single- and
  multi-layer methods. In Fig.~\ref{fig:percolation_diagram_all}, we
  show the percolation diagrams (including also the results already
  presented in Fig.~\ref{fig:fig1}) in duplex networks with no overlap
  (panel a) and with high overlap (panel b).  It is clear that the
  strategies that incorporate multilayer information (i.e., HDA, EMD,
  \DCI, and \DCIz), as well as Pareto-efficient strategy that take one
  of them into account, perform consistently better than those based
  exclusively on single-layer metrics [i.e., ($CI_{\ell=2},
    CI_{\ell=2}$), (CoreHD, CoreHD), and (BPD, BPD)].  This is
  because, as noted in
  Refs.~\cite{Buldyrev_2011_interdependent,osat2017optimal,Baxter2018targeted},
  the presence of interdependencies in the multiplex structure deeply
  affects the overall robustness of duplex networks against random and
  targeted attacks, and this information is not present in either of
  the layers considered separately.}

\newtext{In Fig.~\ref{fig:percolation_vs_overlap_all} we report the
  size of the critical attack set $q$ as a function of structural edge
  overlap. As expected, the best performing targeted strategies for
  $o=1$ are those based on BPD, which is the best-performing strategy
  on single-layer
  graphs~\cite{zdeborova2016fast,Makse_review_2019}. It is interesting
  to notice that some Pareto-efficient strategies outperform Simulated
  Annealing for large values of overlap (same implementation as the
  one presented in~\cite{Baxter2018targeted} with temperature steps
  equal to $10^{-7}$).}

\newtext{Finally, in Fig.~\ref{fig:percolation_vs_rhocorrelation_all}
  we show the size of the critical set found by each of the 20
  strategies for different combinations of interlayer degree
  correlations and edge overlap (same conditions presented in
  Fig.~\ref{fig:fig3}). It is clear that Pareto-efficient strategies
  combining multi- and single-layer information perform better than
  the others also in this case, and especially much better than
  methods relying only on single-layer metrics. This is even more
  evident when the duplex has $o \approx 0$ while the gap becomes
  smaller as the overlap increases, as expected.}

\newtext{
  \section{Tuning inter-layer degree correlation and edge overlap}
  \label{appendix:numerical_methods}
  The algorithm to tune inter-layer degree correlations and edge
  overlap in a duplex network used in the paper is based on biased
  edge rewirings. The procedure works along the same lines of the two
  procedures to separately tune edge overlap and inter-layer degree
  correlations originally considered in
  Refs.~\cite{Nicosia2015,Diakonova2016irreducibility,Santoro_complexity2019}.
  More precisely, to decrease edge overlap we start from two (possibly
  different) layers, we iteratively select at random two edges on a
  randomly chosen layer, and we rewire at random the endpoints of the
  two links only if such rewiring results in a reduction of the edge
  overlap. The procedure is iterated until we reach the desired value
  of edge overlap $o^\star$.  In this way, the degree sequence on each
  layer is preserved throughout the process. Notice that the actual
  range of edge overlap attainable with this method actually depends
  on the degree sequences at the two layers.}

\newtext{Increases in edge overlap are obtained with a similar
  procedure, where a rewiring is accepted only if it results in the
  increase of edge overlap of at least one of the two edges involved
  in the rewiring. As a consequence, also this procedure does not
  modify the degree sequence of each layer. In general, the actual
  range of edge overlap obtained by this procedure depends on the
  actual degree sequences of the two layers. For instance, a value of
  $o=1$ is attainable only if the degree sequences of the two layers
  are identical.  Since the maximum value $o_{\rm max}$ of edge
  overlap for a generic pair of layers is not known a-priori, in our
  simulations we computed an approximation of $o_{\rm max}$ by iteratively
  increasing the overlap of the system until no further increase is
  attainable (i.e. the termination criteria is such that the edge
  overlap does not increase after $5\times 10^7$ random rewirings).}

\newtext{The procedure for tuning the interlayer degree correlation
  $\rho$ is identical to the one presented
  in~\cite{Nicosia2015,Mammult}. Briefly, starting from a generic
  duplex network, we consider $R$ to be the $N \times N$ matrix that
  accounts for the coupling between the nodes of the two layers. Here,
  the generic entry $r_{ij}=1$ if node $i$ in layer $\alpha$
  corresponds to node $j$ in layer $\beta$.  Since we are dealing with
  a duplex network, there is a one-to-one correspondence between the
  nodes in the two layers, so that we have to impose $\sum_j
  r_{ij}=1\; \forall i$. The main idea is that the coupling $R$ can be
  realised in many ways, and among all these possibilities we choose
  one of those that correspond to a given level of interlayer degree
  correlation $\rho^\star$. We define the cost function $F(R) =
  |\rho-\rho^\star|$, and we iteratively modify the structure of the
  matrix assignment in order to minimize $F(R)$.  The minimisation
  procedure is obtained by using a simulated annealing algorithm. In
  particular, starting from a certain matrix assignment $R$, we rewire
  two edges at random of such matrix in order to obtain a new
  assignment $R'$. We then accept the new assignment with
  probability:
  \begin{equation*}
    p = 
    \begin{cases}
      1 & \text{ if } F(R') < F(R) \\ {\rm e}^{-\frac{F(R') - F(R)}{\beta}}
      & \text{otherwise}
    \end{cases}
  \end{equation*}
  where $\beta$ has the role of an inverse temperature. The algorithm
  stops when $F(R) < \varepsilon$, where $\varepsilon$ is a threshold
  set by the user. In our simulations, we consider $\beta=10^{-7}$ and
  $\varepsilon=0.005$.}
\newtext{The two algorithms to tune edge overlap and inter-layer
  degree correlations can be combined to obtain a duplex network with
  prescribed values of $o$ and $\rho$. We start by tuning the value of
  inter-layer degree correlation $\rho$, and we then iteratively
  increase edge overlap $o$ through biased edge rewiring. Notice that
  the combination of these two procedures (in this order) does not
  alter the original degree distribution on each layer of the duplex
  network.}

\bibliographystyle{unsrt}

\end{document}